\documentclass[12pt]{amsart}
\usepackage{times}
\usepackage{overcite}
\usepackage{url}
\usepackage{revsymb}
\usepackage{graphicx}
\usepackage{titlesec}
\usepackage{subfigure}
\usepackage{float}
\usepackage[footnotesize]{caption}
\usepackage[pdftex,colorlinks=true,urlcolor=blue,citecolor=blue,linktocpage=true]{hyperref}

\titleformat{\section}[hang]
{\normalsize\scshape\centering} 
{\thesection.}    
{.3em}            
{}                
\titlespacing*{\section} 
{0pt}    
{*0.7}  
{*0.4}  
\titleformat{\subsection}[hang]
{\large\bfseries} 
{\thesubsection.}    
{.5em}            
{}                
\titlespacing*{\subsection} 
{0pt}    
{*0.4}  
{*0.2}  
\titleformat{\subsubsection}[hang]
{\normalsize\bfseries} 
{\thesubsubsection.}    
{.5em}            
{}                
\titlespacing*{\subsubsection} 
{0pt}    
{*1.0}  
{*0.2}  
\titleformat{\paragraph}[runin]
{\normalsize\bfseries} 
{}                
{0pt}             
{}                
[~~]             
\titlespacing*{\paragraph} 
{0pt}    
{*0.2}   
{*0.2}   

\def\go{\mathrel{\raise.3ex\hbox{$>$}\mkern-14mu\lower0.6ex\hbox{$\sim$}}}
\def\lo{\mathrel{\raise.3ex\hbox{$<$}\mkern-14mu\lower0.6ex\hbox{$\sim$}}}

\renewenvironment{itemize}
{\begin{list}{$\bullet$}
        {\setlength{\topsep}{0pt}
         \setlength{\itemsep}{0pt}
         \setlength{\parsep}{0.25\parsep}
         \settowidth{\labelwidth}{$\bullet$}
         \setlength{\itemindent}{0pt}
         \setlength{\leftmargin}{0pt}
         \addtolength{\leftmargin}{\labelwidth}
         \addtolength{\leftmargin}{\labelsep}
}
}
{\end{list}}

\makeatletter
\newcommand\fs@boxedcaptioninside{\def\@fs@cfont{\bfseries}\let\@fs@capt\floatc@plain
  \def\@fs@pre{\setbox\@currbox\vbox{\hbadness10000
    \moveleft3.4pt\vbox{\advance\hsize by6.8pt
      \hrule \hbox to\hsize{\vrule\kern3pt
        \vbox{\kern3pt\box\@floatcapt\kern5pt\box\@currbox\kern3pt}\kern3pt\vrule}\hrule}}}%
  \def\@fs@mid{\kern2pt}%
  \def\@fs@post{}\let\@fs@iftopcapt\iffalse}
\makeatother

\floatstyle{boxedcaptioninside}
\newfloat{Box}{t}{box}[section]

\def\footnote#1{\ifthenelse{\value{footnote}=9}{\setcounter{footnote}{0}}{}%
\footnotemark\footnotetext{#1}}

 \author[Nelemans et al.]{ G. Nelemans$^1$ \and M. Wood$^{2,1}$ \and
   P. Groot$^1$ \and S. Anderson$^3$ \and K. Belczynski$^4$ \and
   M. Benacquista$^5$ \and P. Charles$^6$ \and A. Cumming$^7$ \and
   C. Deloye$^8$ \and P. Jonker$^{9,10}$ \and V. Kalogera$^8$ \and
   C. Knigge$^{11}$ \and T. Marsh$^{12}$ \and P. Motl$^{13}$ \and
   R. Napiwotzki$^{14}$ \and K. O'Brien$^{15}$ \and E.S. Phinney$^{16}$ \and
   G. Ramsay$^{17}$ \and T. Shahbaz$^{18}$ \and J.-E. Solheim$^{19}$ \and
   D. Steeghs$^{12}$ \and M. van der Sluys$^{8}$ \and F. Verbunt$^{20}$ \and
   B. Warner$^{21}$ \and K. Werner$^{22}$ \and K. Wu$^{23}$ \and
   L. R. Yungelson$^{24}$}

 \title[Ultra-compact binaries]{The astrophysics of
   \emph{ultra-compact binaries}\\ \smallskip {\small A white paper
   for the Astro2010 decadal review}}

\begin{document}

\maketitle

\bigskip 
\noindent {\small
1 Radboud University Nijmegen, The Netherlands\\ 
2 Florida Institute of Technology, Melbourne, FL\\
3 University of Washington, Seattle, WA\\
4 Los Alamos National Laboratory, NM\\
5 University of Texas at Brownsville, TX\\
6 South African Astronomical Observatory, South Africa\\
7 McGill University, Montreal, Canada\\
8 Northwestern University, Evanston, IL\\
9 Harvard Center for Astrophysics, Cambridge, MA\\
10 SRON, Netherlands Institute for Space Research, Utrecht, The Netherlands\\
11 University of Southampton, UK\\
12 University of Warwick, UK\\
13 Indiana University Kokomo, IN\\
14 University of Hertfordshire, UK\\
15 European Southern Observatory, Chile\\
16 Caltech, Pasadena, CA\\
17 Armagh Observatory, Northern Ireland, UK\\
18 Instituto de Astrofisica de Canarias, Spain\\
19 University of Oslo, Norway\\
20 University of Utrecht, the Netherlands\\
21 University of Cape Town, South Africa\\
22 University of T\"ubingen, Germany\\
23 Mullard Space Science Laboratory, University College London, UK\\
24 Institute of Astronomy of the Russian Academy of Sciences, Moscow, Russia\\
}

\bigskip
\bigskip

\textbf{Primary panel: SSE}\\

\textbf{Secondary panels: GAN, CFP}\\

\newpage

\noindent \textbf{Definition}\\ \textit{Ultra-compact binaries are objects
  which have orbital periods shorter than one hour.  Both stars must be
  compact and are typically degenerate and hydrogen deficient.  The class
  includes interacting AM CVn stars, ultra-compact X-ray binaries, detached
  double white dwarfs, double neutron stars, white dwarf/neutron star binaries
  and as yet unobserved binaries such as black holes with neutron star or
  white dwarf companions.}

\medskip
\noindent \textbf{Scope}\\ Ultra-compact binaries allow us to study unique
(astro)physical phenomena specific to their ultra-short periods and the
H-deficient nature of the components. This white paper focuses
on:\\ 1.\ Gaining insight in binary evolution by understanding the formation
of these systems. 2.\ Understanding fundamental physics of gravitational wave
emission and mass transfer via observation and theory. 3.\ The question of
mass-transfer stability and the (often explosive) physics of compact object
mergers.  4.\ Charting the ultra-compact binary population.  5.\ Testing
accretion physics in H-deficient regimes.

\section*{Astrophysical context and Relevance}

Stars are the building blocks of the Universe, and understanding the physics
of their formation and evolution is foundational to making progress in many
branches of astrophysics. Binary stars play a special role in this for several
reasons. Interacting binaries allow us to probe stellar evolution processes
that cannot be studied in single stars, such as literally exposing the
internal structure of stars. They also give rise to energetic,
accretion-driven phenomena, e.g.\ type Ia supernovae (SNIa), that can be used
to identify and study stellar populations out to cosmological distances. Only
in binary systems can stellar masses be measured accurately and progenitor
masses inferred, constraining the major open question of the death of stars:
which stars form white dwarfs (WD), neutron stars (NS), and black holes (BH),
and how does this take place?  Finally, the large fraction of stars in binary
systems makes the study of binaries essential for understanding of the physics
of stellar evolution.

Ultra-compact binaries are particularly relevant for the question of \emph{the
  formation of compact binaries}, which requires the formation of compact
stellar remnants as well as significant angular momentum loss via
common-envelope stages. This process creates the vast majority of objects
responsible for high-energy phenomena in the Universe. Ultra-compact binaries
have often experienced two such episodes.  They are also spectacular
\emph{probes of the extreme physics} at high energy and high density, being
strong gravitational wave (GW) sources (the only guaranteed sources for LISA,
Fig.~1). They harbor accreting millisecond pulsars and when merging they may
explode as SNIa or short gamma-ray bursts (GRBs), leave low mass WDs or
hydrogen-deficient giants.

Ultra-compact binaries are of general interest as laboratories for
studying \emph{the physics of accretion}, which produces the bulk of
high-energy radiation in the Universe and is central to understanding such
disparate phenomenon as SNIa explosions and the growth of super-massive
BHs. Ultra-compact binaries are special since they combine the accretion of
hydrogen deficient material (a feature unique to this population) with short
time scales for the accretion processes. This offers the possibility of
testing the physical theories in a substantially different environment.

Progress in studying ultra-compact populations is currently hampered by the
small number of systems known.  However, the next decade holds enormous
promise for the study of ultra-compact binaries, due to the advent of large
wide-field (and variability) surveys to find new systems, efficient optical and
near-IR spectrographs combined with (extremely) large telescopes to allow
detailed characterization of their properties, and the imminent advent of GW
astronomy, which will allow studying these populations in unprecedented
detail.

\section{Gaining insight in binary evolution: How do ultra-compact binaries form?}\label{formation}

An ultra-compact binary consists of compact stars, formed from the cores of
well-evolved stars, and thus has much lower orbital energy and angular
momentum than the progenitor binary that contained giants[1]. The binary is
thought to shrink mainly during a phase of unstable mass transfer and
ejection, the spiral-in. If the outcome of this process is derived by assuming
that the change in orbital energy is enough to eject the giant's mantle, the
predicted properties of the ultra-compact binary do not match the observations
of double white dwarf binaries.  These properties can be matched with the
assumption that the giant's mantle is ejected carrying the specific orbital
angular momentum[2], but this begs the question how the required energy is
provided. It is clear that a more complete theoretical description is required
that takes into account both energy and angular momentum. For neutron stars
and black holes the existence and magnitude of asymmetric kicks imparted
during the supernova event add to the uncertainty[3]. Of special interest is
the formation of ultra-compact binaries in stellar clusters, both because many
field systems must have originated in clusters and because ultra-compact
systems are over abundant and seem to have different properties in dense
clusters, implying that dynamical encounters play a role in their origin[4].

To proceed, we envisage a two-pronged approach using detailed studies of
individual systems on one hand, and of the whole population on the other
hand. Any viable evolution scheme must be able to reproduce the exact
properties (such as component masses, orbital period, age, system velocity) of
each observed system. Such scheme must also reproduce the distributions of and
correlations between these properties in the population of ultra-compacts.
Observationally this implies the accumulation of large homogeneous samples of
ultra-compact binaries, and the detailed follow-up of a number of individual
systems.

\noindent \textbf{ Prospects}: Significant progress can be expected in the
next decade, in particular from the combination of the large-scale (X-ray, UV,
optical, NIR, radio and GW) surveys paired with population modeling to explore
the full parameter space of possible binary evolution. When combined, the
convolved uncertainties in binary evolution will be significantly
constrained. Much more realistic physical models for star formation,
supernovae and even common envelopes will be possible. Improved numerical
modeling of dense stellar systems combined with deep observations of globular
clusters will constrain dynamical formation processes.

\section{Understanding physical interactions: evolution and mass transfer}\label{evolution}

Once formed as a detached ultra-compact binary, the system will evolve towards
shorter orbital period via angular momentum loss until the components reach
contact. The system merges (Sect.~\ref{mergers}) or survives as an
\emph{interacting} binary, evolving to longer periods as a result of mass
transfer, even though the orbital angular momentum still decreases. Known
interacting systems are the double WDs or AM CVn systems and WD-NS systems
(ultra-compact X-ray binaries, UCXBs). The UCXBs are strongly overproduced in
globular clusters and may actually be the dominant population of X-ray
binaries[7].

For the orbital shrinking phase, GW radiation and possibly angular momentum
loss via magnetic stellar winds dominate, but tidal effects and interactions
between magnetic components can alter the evolution[8]. Associated time scales
in these ultra-compact binaries are short enough to measure period evolution
on human time scales, in particular for eclipsing systems (as found
recently[9]), systems harboring radio pulsars and when GW measurements become
available (which can probe even $\ddot{P}_{\rm orbit}$).  This allows tests of
GW predictions as well as magnetic interactions and gives a handle on the
importance of tides in these ultra-dense objects.

In the expanding phase, mass transfer/loss and associated angular momentum
loss complicate the orbital evolution. Still, time scales are short and the
evolution can be followed in real time by measuring period derivatives, while
mass-transfer rates can be inferred from the (bolometric) luminosity and
system parameters when (parallax) distances are known[10]. The mass-transfer
rate is set by a combination of the GW losses and the internal structure of
the donor stars, which thus can be inferred, although magnetic interaction and
self-irradiation need to be taken into account. The donor stars have, so far,
never been observed, but when irradiated they may be detectable in the
IR. Abundance patterns in the transferred material can be determined from
accretion disk spectra (Fig.~1) and often can be used to determine the
progenitor system[11]. Due to the compact size of the systems, the disks
hardly emit in the (near-)IR, allowing study of the unknown mechanism of jet
launching from compact objects, in that wavelength range[12].

\begin{figure}
\hspace*{-0.5cm}
\includegraphics[angle=-90,width=0.5\textwidth]{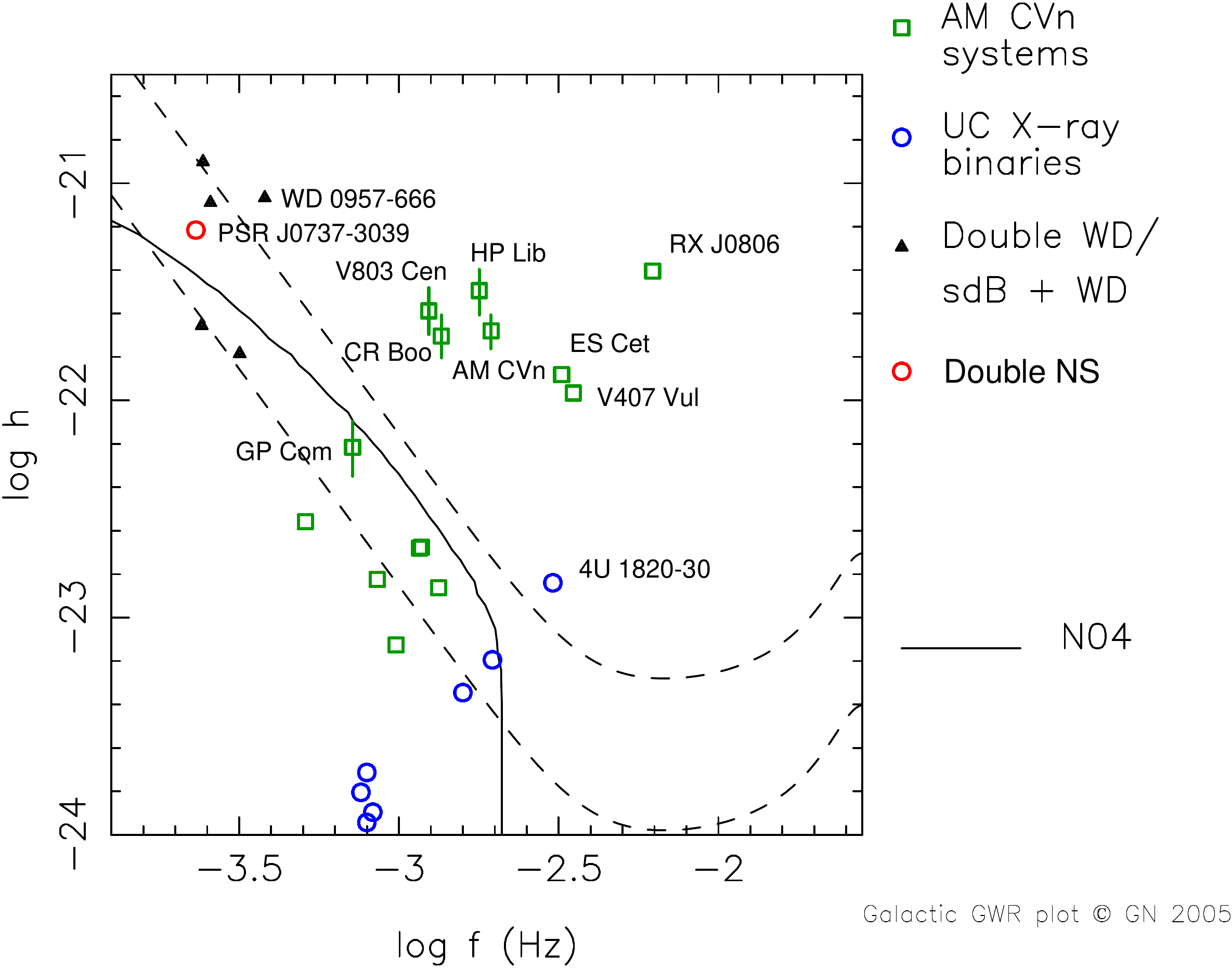}
\includegraphics[angle=-90,width=0.45\textwidth]{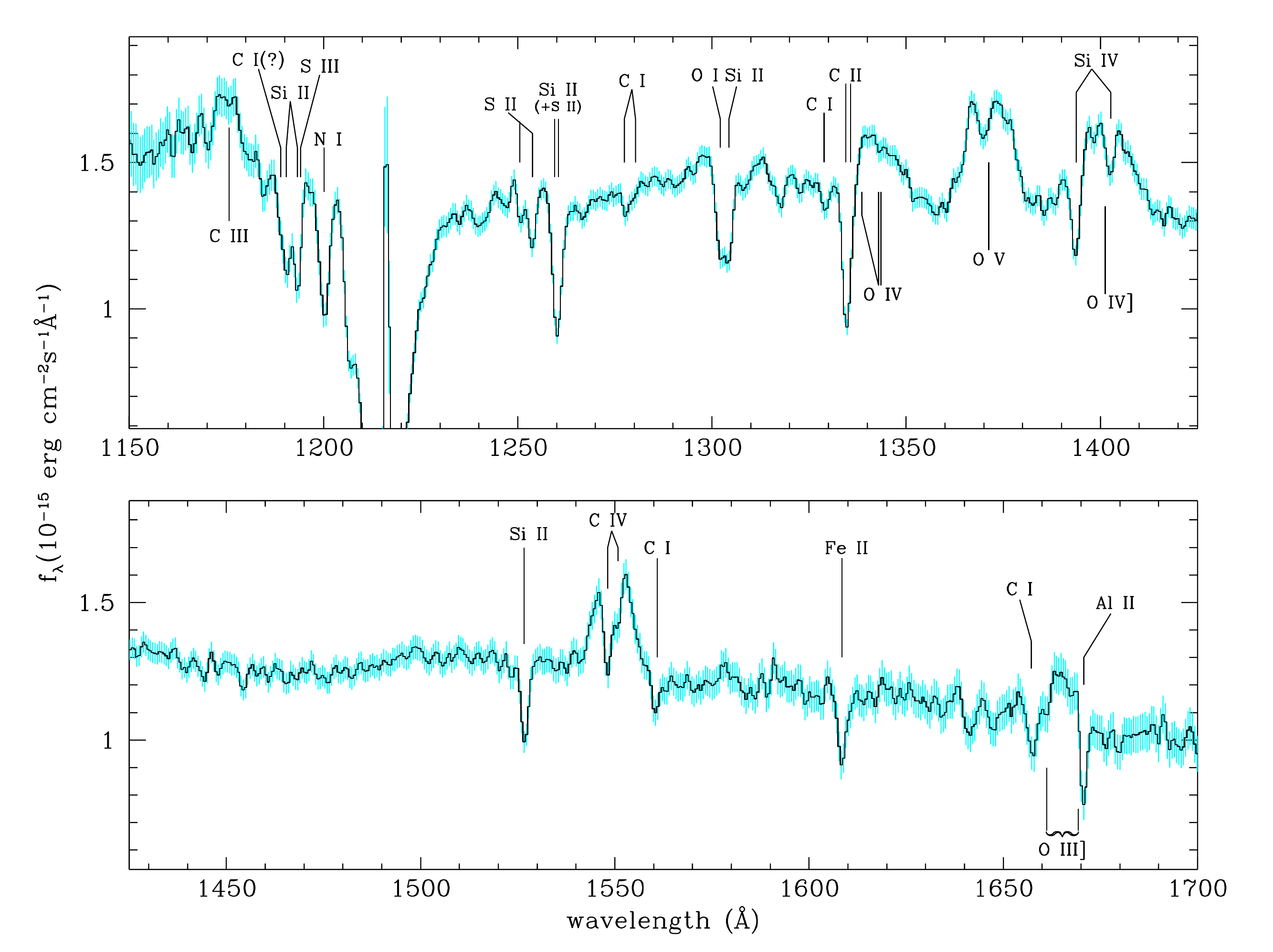} 
\caption[]{Left: known ultra-compact binaries as GW sources (frequency
  vs.\ strain) compared to the LISA sensitivity. Right: UV spectrum of the UCXB
4U 1626-67 [5]}
\end{figure}

\noindent \textbf{Prospects}: With dedicated, short timescale variability
surveys and piggy-back programs to find many new systems
(Sect.~\ref{numbers}), the prospects of using eclipsing systems for
evolutionary studies are excellent.  Direct measurement of the evolution of
the shortest period interacting systems (via GW measurements or X-ray, UV,
optical and radio monitoring of the newly discovered systems), combined with
determination of current mass-transfer rates for the longer period systems,
will provide stringent tests of the accepted GW driven evolution model. For
this distance measurements are crucial (VLBI, GAIA). The results of detailed
nucleosynthesis calculations combined with binary evolution scenarios can be
compared with high quality (UV/optical/NIR) spectra to test the formation
mechanisms.

\smallskip
\section{When is mass transfer stable and what happens in compact object mergers?}\label{mergers}

A critical juncture in the life of an ultra-compact binary is the period just
after mass transfer commences in the system, called contact (which occurs at
orbital periods between 2 and 15 minutes for systems with at least one WD, and
at millisecond periods for NS and BH systems). The system either merges, or --
when at leat one WD is present -- may survive to become an interacting binary.

To survive, the mass transfer must be stable, and this depends on several
factors: the mass ratio, angular momentum loss mechanisms, and the internal
structure of the donor[13]. Accretion in these systems can proceed directly
onto the accretor[14], which destabilizes mass-transfer by sinking angular
momentum into the accretor's spin.  This so-called ``direct impact'' phase
(Fig.~2) may very well be associated with two very-short period binaries found
as ROSAT X-ray sources, HM Cnc and V407 Vul (the shortest period binaries
known at 5.4 and 9.5 min resp.). Tidal coupling can return angular momentum to
the orbit (but on very uncertain time scales[13]), heating both components at
the same time. This heating can influence the donor's response to mass loss
and hence affect mass transfer stability. Even if mass transfer is stable,
binary survival is not guaranteed.  Mass transfer rates can become
super-Eddington, and it is unclear what happens in these cases.
Accretion-sparked thermonuclear explosions on the accretor may also destroy
the binary during this phase shortly after first contact
(Sect.~\ref{accretion}). The properties of the surviving population depend
critically on the interplay of these physical processes and can be used to
constrain them.

When mass transfer is unstable, a catastrophic merger event cannot be
avoided. The further evolution of the rapidly rotating merger product can
produce several extremely interesting phenomena, including: collapse to a NS,
formation of an R CrB star (helium-rich giant), or explosion as a SNIa. In the
context of NS and BH binaries, mergers can lead to short GRBs. The resulting
disruption of a NS in such merger events sheds light on the nuclear equation
of state and physics at extreme density. While theoretical advances detailing
the possible outcomes and the phenomenology of the mergers have been made in
the last decade (with significant breakthroughs in the fully relativistic
modeling of BH mergers[15]), basic questions remain: Can WD mergers explode as
SNIa?  If so, under what conditions? Do merging NSs and BHs produce GRBs?

\noindent \textbf{Prospects}: The wide field and variability surveys will find
many new ultra-compact binaries and extra-galactic merger events. Wide-field
spectroscopic studies can be used for confirmation of candidates. Larger
samples will enable us to determine relative numbers of shrinking and
expanding systems and thus to directly determine the survival and merger
fractions. A combination of increased numbers of observed SNIa and GRBs and
better statistics on their occurrence in different environments will provide
the observational handle on the question whether they are (sometimes) related
to mergers. This should be complemented by further theoretical modeling of the
merger processes and the early contact phase of WD ultra-compact binaries. The
surviving binaries must be studied in detail to determine the distribution of
their properties, which constrain the physics influencing survival. Finally,
GW studies of the NS and BH mergers with high-frequency detectors such as LIGO
and VIRGO, and the observations of WD mergers and pre and post-contact WD
systems with LISA will provide unprecedented detail (LIGO) and statistics
(LISA).

\section{Charting the ultra-compact binary population}\label{numbers}

Many of the issues discussed earlier have a strong influence on the number of
ultra-compact binaries that exist in our Galaxy. Conversely, determining the
absolute and relative number of `flavors' of ultra-compact binaries provide
direct tests of the physics of their formation, evolution, mass-transfer
stability and merging.

Current predictions based on population modeling (Sect.~\ref{formation}),
combined with calibration from the observed systems[16], estimate the Galactic
numbers at around 10$^8$ for double WDs, several 10$^6$ each for double NS-NS
and NS-WD systems. A large population of ultra-compact binaries with NS and/or
BH with periods below 1 hr may form, strongly influencing merger
rates[17]. The predicted numbers of AM CVn systems and UCXBs are very
uncertain at around 10$^6$ for each case[18]. Most of the studies up to now
have concentrated on particular populations, but the relative numbers are
easier to determine and will give insight in the relative importance of
different formation channels.  The rates at which the systems get into contact
is of particular interest, as this may lead to (explosive) mergers or to the
formation of interacting binaries (Sect.~\ref{mergers}).  A combination of
finding more systems before and after they merge (for which recognizing the
merger product(s) is a pre-requisite) has a great potential for constraining
merger rates.

\noindent \textbf{Prospects}: The next decade promises to be a golden age for
the discovery of new systems, in particular large samples with homogeneous
selection effects due to the advent of large-scale, wide-field surveys. GW
discoveries with LISA, will identify the spectral shape of the unresolved
fore-
\noindent ground and more than 10,000 individual systems (complete for periods
shorter than $\sim$12 min!). These samples will shed light on binary evolution
models and the star formation history of the Galaxy. Distance measurements, in
particular by GAIA and LISA (for several thousand), give the necessary third
dimension to determine space densities Galactic structure. The (planned)
increase in the scale of the variability surveys (e.g.\ Palomar Transient
Factory, Pan-Starrs, LSST) will open up the possibility of directly observing
even very rare events such as mergers and common-envelope phases. For the
brighter sources and events, extra-galactic studies will become an important
new strand of investigation, largely removing the problems associated with the
uncertain distances for the Galactic populations.

\begin{figure}
\includegraphics[width=0.45\textwidth]{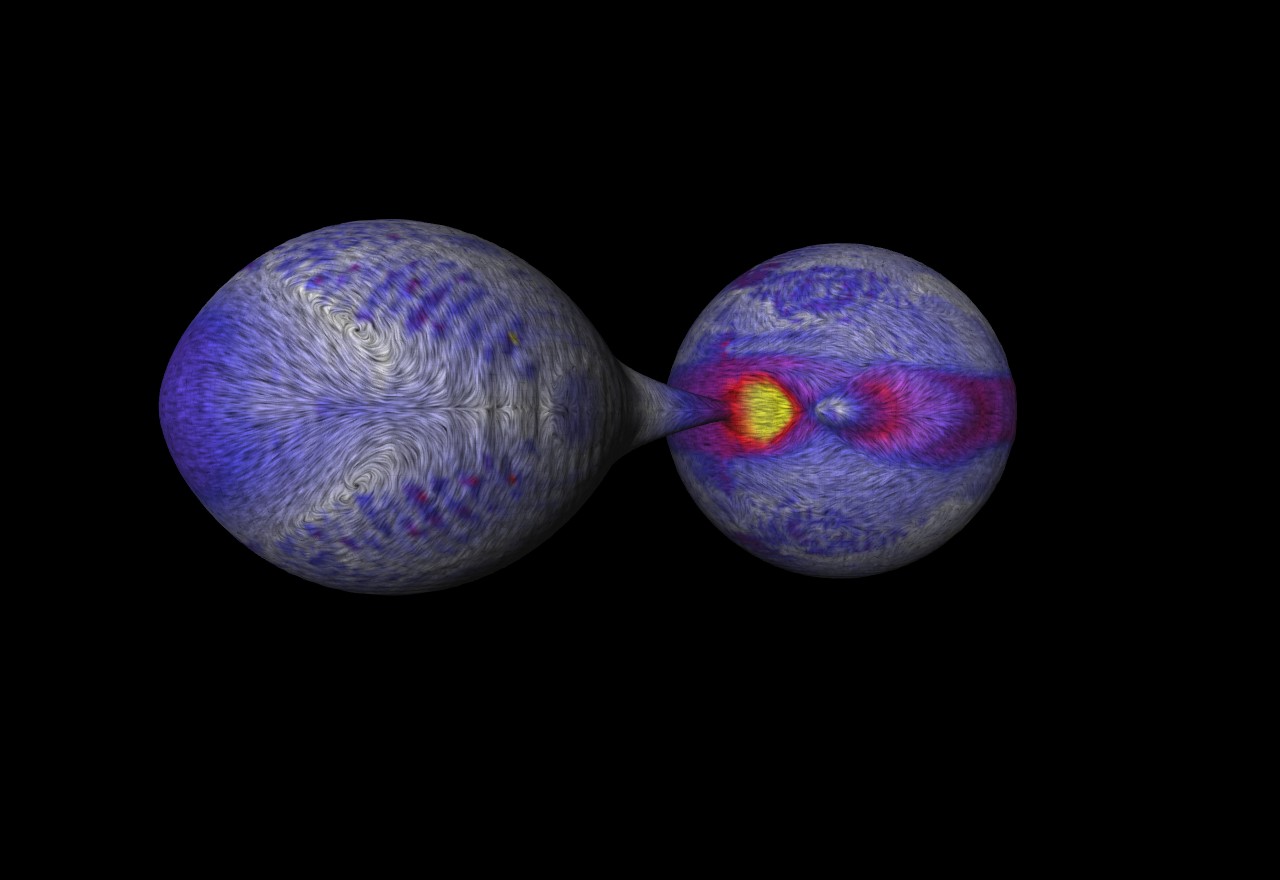}
\includegraphics[width=0.45\textwidth]{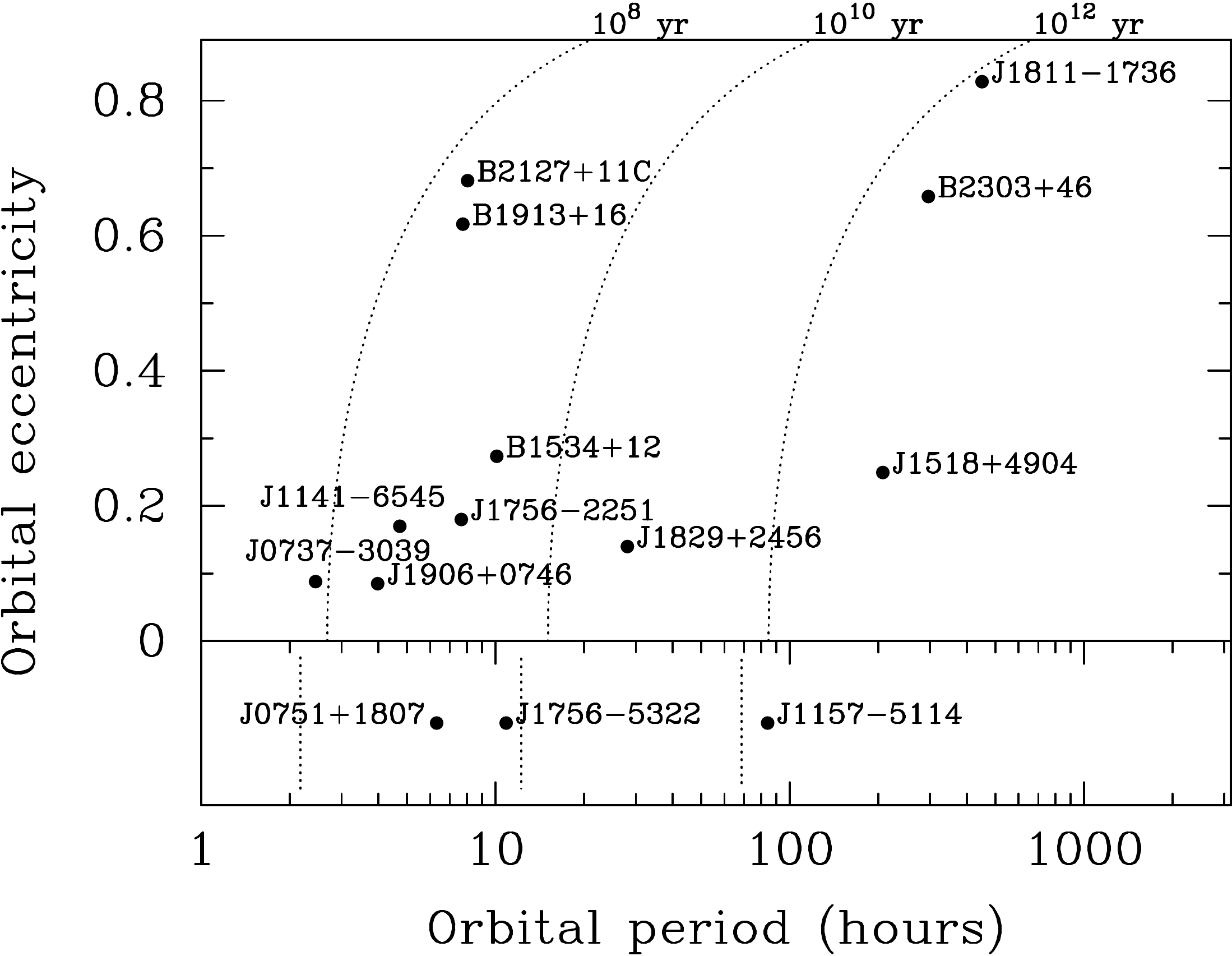} 
\caption[]{Left: Snapshot of a simulation of direct impact accretion (Motl,
  priv. comm). Right: period -- eccentricity of the known NS binaries with
  their merger time indicated[27]. }
\end{figure}

\section{Testing the physics of accretion in H deficient situations}\label{accretion}

Because in interacting ultra-compact binaries the transferred matter is H
deficient (typically He or C/O rich, e.g. Fig.~1), accretion disk theories
developed for H-rich gas can be tested in different environments. For instance,
the disk-instability model applied to He disks suggests the possibility of
steady cool (i.e.\ neutral) disks[19], which indeed seem to be observed in the
long-period AM CVn systems, while applied to UCXBs the disk-instability model
seems to fail to explain some persistent systems[20]. The observed He and C/O
disk spectra are very different from their H-rich counterparts. For instance
the He-rich disks in AM CVn stars show a mysterious ``central spike,'' a third
emission peak (in addition to the omnipresent double peaks from the disk) that
seems to originate from the accretor, which has never been observed in H-rich
systems. There is considerable progress in the (spectral) theoretical models
for H-deficient accretion disks, but they are not yet accurate enough for
quantitative comparison with the observations[21].

The H-deficient material ends up on the accreting WD or NS. Accumulating
H-rich material on WDs leads to expansion to giant dimensions, steady H
burning or shell flashes (novae), depending on the specific accretion rate and
the local gravity. For He accretion the overall scheme is similar, although
much work is still needed to determine the exact picture. The big difference
is that for the He ignition much higher pressure and densities are needed,
leading to much more violent and rare He novae, as compared to H novae.  At the
relatively low accretion rates reached in AM CVn systems the ignitions may be
so violent that they lead to thermonuclear explosions, somewhat like small
scale SNIa (termed .Ia supernovae[22]). He accretion may also alleviate some
of the problems that may prevent H accreting WDs from reaching the
Chandrasekhar mass and exploding as full-blown SNIa[23].  To date no C/O
accreting WDs have been observed and their fate has not been studied in
detail.

Accretion onto a NS often leads to thermonuclear explosions that are observed
as type I X-ray bursts. The situation is slightly different from WD accretors,
as often in H-rich accretion the H is first (in part) stably burnt into He,
which then ignites unstably, or even both H and He burn stably into a C layer
that then explodes in a so-called super burst[24]. The difference between
H-rich and He- or C/O-rich accretion then becomes more subtle, as the
He-bursts from steadily burnt H are different from the He bursts from He
accretion, only due to the lower temperature in the accumulated He layer in
the latter case. Although from optical, UV and X-ray spectra there is evidence
for accretion disks that are C/O dominated and thus He deficient, no C/O
bursts have been observed. In contrast, for some sources that seem to have C/O
disks, bursts have been observed that are best explained as He bursts[25]!
Apart from obvious solutions such as misinterpretation of either burst or disk
spectra, one possibility would be the spallation of C/O nuclei into He nuclei,
a spectacular nuclear physics experiment.

\noindent \textbf{Prospects}: With the expected significant enlargement of the
number of known systems, the prospects are excellent for obtaining spectra of
many H-deficient accretion disks, to probe composition, structure, winds and
jets, over a large range of parameters. Progress in theoretical understanding
of accretion disk spectra can be anticipated because of recent developments in
(magneto)hydro-dynamics calculations[26]. A systematic study of accretion
disks and their stability for different chemical compositions -- different
molecular weights, ionization temperatures, opacities and viscosities -- may
become an important ingredient for answering the now 30-year-old question of
the structure and stability of accretion disks.

The many newly discovered systems will allow systematic study of disk
instability outbursts and the (thermonuclear) explosions on WDs and NSs.
Combining the novae/burst observations with the chemical composition
information coming from the disk spectra will allow an unprecedented detailed
test of our understanding of accretion and ignition physics, and thus provide
a solid basis for understanding in which circumstances accretion onto WDs may
push them over the Chandrasekhar limit, so that they explode as SNIa.

\section*{Necessary Facilities}

\begin{itemize}
\item \textit{Wide-field and variability surveys} For optical work many
  projects are under way or planned, (e.g.\ SDSS, PTF, Pan-Starrs, SkyMapper,
  ESO VST, Vista, LSST). Limited sampling in the UV is done by GALEX,
  XMM-Newton, Swift, TAUVEX so new missions are needed (e.g.\ the Russian
  WSO-UV; HORUS and later MUST, ATLAST). Small-scale X-ray surveys are being
  undertaken using Chandra and XMM-Newton so new missions are needed (many are
  proposed, of which the eRosita instrument is most advanced).  Radio surveys
  are needed to find new pulsars. GW measurements with LISA will discover
  thousands of new systems and LISA and LIGO will provide unique information
  on the mergers and their rates.
\item \textit{X-ray/UV/Opcital/NIR spectrographs} Wide-field/multi-object
  spectroscopic instruments (such as WFMOS) are needed for
  confirmation/follow-up of candidates. For detailed (kinematics, abundance)
  studies, X-ray, UV, optical and NIR spectrographs with sufficient resolution
  (R several thousand) and throughput (e.g.\ X-Shooter), mounted on large to
  very large telescopes, are needed (with magnitudes of 21-22 even 8-m-class
  telescopes are often too small, in particular when phase resolved
  spectroscopy is needed). High-speed (photon counting) detectors are needed
  to cover the shortest time scales.  For calibration of luminosities,
  parallax measurements with GAIA or VLBI are crucial.
\item \textit{Monitoring and high-speed instruments} All-sky X-ray monitors
  are important for finding variable objects. Eclipsing systems, found by
  GAIA, LSST and LISA must be monitored with optical, UV, X-ray and radio
  instruments to follow the long(er) term evolution. For that it is important
  to keep high-speed instruments on modest-sized facilities (such as an
  international network of dedicated (small) telescopes, e.g\ Las Cumbres
  Observatory and SONG) and to keep high time resolution in future UV
  missions.  GWs as measured by LISA will directly follow the evolution of
  thousands of the shortest period systems. Monitoring of radio pulsars in
  binaries will reveal the evolution of NS systems.
\item \textit{Theoretical and numerical tools} Detailed theoretical modeling
  of the evolution of binaries, the onset of mass transfer and the merger
  process itself relies on further developing (relativistic) numerical methods
  (combining hydrodynamics with radiation and magnetic interaction) and
  continued developments of computer power.  Significant further efforts in
  theoretical spectral modeling of accretion disks, which requires
  understanding of their physical structure are needed.
\end{itemize}

\section*{Conclusion}

We conclude that a large number of ultra-compact binaries, studied in great
detail, will allow us to study a number of (astro)physical phenomena that are
of general importance in our understanding of the Universe, in particular the
formation of all compact binaries, many high-energy phenomena such as SNIa and
GRBs, and our general understanding of accretion. In the next decade this
possibility will become reality as there is enormous promise for finding new
systems, in particular with optical wide-field and variability surveys and
even more with LISA.

These large numbers will enable statistical studies that constrain binary
evolution, but also fundamental physics of interaction and merger. This can
only be done if Galactic studies of the faint and common systems are combined
with studies of the rare (but explosive) phenomena far away.

\smallskip
\noindent \textbf{Recommendations} Crucial facilities are
\begin{itemize}
\item GW detectors (LISA, LIGO) for finding and following
  systems up to merger.
\item Wide field and variability surveys (PTF, Pan-Starrs, SkyMapper, VST,
  Vista, LSST) to find (rare) systems and phenomena.
\item X-ray instruments (timing and spectra, as proposed by many notices of
  interest).
\item New large UV spectrograph in space (e.g.\ WSO-UV, HORUS, MUST,
  ATLAST). The lack of missions on the horizon is worrying.
\item Optical (multi-object) spectrographs with R = 5,000-10,000 on
  (extremely) large telescopes (WFMOS, X-Shooter).
\item Radio surveys for detecting pulsar binaries and following their
  evolution.
\item Network of small(er) (X-ray/UV/optical), rapid-responde telescopes
  obtaining light-curves on short (second to hour) as well as long (days to
  weeks) time scales.
\item Sufficient opportunities for obtaining grants for theoretical/numerical
  developments.
\end{itemize}

\smallskip
\noindent \textbf{References}\\ {\small 1 Paczynski, 1976 in Eggleton et
  al. p75; 2 Nelemans et al.\ 2000 A\&A 360 1011, Van der Sluys et al.\ 2005
  A\&A 440 973; 3 Kalogera 1996 ApJ 471 352; 4 e.g.\ Verbunt 1987 ApJ 312 23;
  5 Homer et al.\ 2002 AJ 124 3348; 7 e.g\ Deloye, Bildsten 2004 ApJ 607 119,
  Ivanova et al.\ 2008 MNRAS 386 553; 8 e.g\ Racine et al.\ 2007 MNRAS 380
  381, Wu et al.\ 2002 MNRAS 331 221, Willems et al.\ 2008 PhRvL 1102; 9
  Anderson et al.\ 2005 AJ 130 2230; 10 Roelofs et al.\ 2007, ApJ 666 1174; 11
  Nelemans, Tout 2003 whdw.conf 359, Nelemans et al.\ in prep; 12 Migliari et
  al.\ 2006 ApJ 643 L41; 13 Marsh et al.\ 2004 MNRAS 350 113, Deloye et
  al.\ 2007 MNRAS 381 525; 14 Marsh, Steeghs 2002 MNRAS 331 7; 15 Pretorius
  2005 PRL 95 12110; 16 e.g.\ Maxted,Marsh 1999 MNRAS 307 122, Kim et
  al.\ 2005 ASPC 328 83; 17 Kalogera et al.\ astro-ph/0612144; 18
  Nelemans, astro-ph/0310800, Roelofs et al.\ 2007 MNRAS 382 685; 19 Tsugawa,
  Osaki, 1997 PASJ 49 75; 20 Lasota et al.\ 2008 A\&A 486 523; 21 Werner et
  al.\ 2006, A\&A 450 725; 22 Bildsten et al.\ 2007 ApJ 662 95; 23 e.g. Yoon,
  Langer, 2003 A\&A 412 53; 24 e.g. Cornelisse et al.\ 2000 A\&A 357 21; 25 in
  't Zand et al.\ 2005 A\&A 441 675; 26 Balbus 2003 ARA\&A 41 555; 27 Lorimer
  2005 LRR 8 7}

\end{document}